\documentclass[reprint,aps,pra,floatfix]{revtex4-2}
\usepackage{packages}
\makeatletter
\let\@bibdataout@rev\@empty
\makeatother

\begin{document}

\title{Pauli-string grouping for VQE measurement reduction on a sparse-connectivity quantum annealer}



\author{Raul Martinez}
    \affiliation{Quantum Technologies Department, ITECAM}
\author{Miguel Sánchez-Beato}
    \affiliation{Quantum Technologies Department, ITECAM}
\author{Mario Calonge}
    \affiliation{Quantum Technologies Department, ITECAM}
    

\begin{abstract}
The Variational Quantum Eigensolver (VQE) requires a large number of measurements to evaluate molecular Hamiltonians. Expressing a molecular Hamiltonian as a linear combination of Pauli strings creates a measurement bottleneck: non-commuting Pauli strings cannot be measured simultaneously. Consequently, mutually commuting Pauli strings must be grouped and measured together to minimise the number of quantum-state preparations. This task maps to the minimum clique cover problem on a commutativity graph, an NP-hard problem typically addressed using classical heuristics. Although an Ising-model formulation has recently been explored on fully connected CMOS Ising machines and demonstrated on physical quantum annealers only at small scale, the embedding cost that governs its behaviour on hardware with sparse connectivity, where each logical variable must be represented by a chain of physical qubits, has not been characterised. In this work, we formulate the Pauli-grouping problem as a standard QUBO colouring model and study its scalability on a D-Wave quantum annealer. Across a series of molecular systems and for both qubit-wise and full commutativity, we quantify the growth in the number of logical variables and the QUBO interaction density, characterise the physical-qubit and chain-length overhead required for embedding on the Zephyr topology, and compare the annealer's time-to-solution and solution quality with those of heuristic classical baselines. This analysis identifies the threshold of molecular complexity beyond which hardware connectivity prevents viable embedding, and shows that a second, practical limit is reached earlier. The threshold therefore measures how far current annealers are from the regime in which pre-optimising a VQE measurement scheme on hardware would be worth considering.

\end{abstract}

\maketitle

%

\section{Introduction} \label{sec:Introduction}
The simulation of molecular electronic structure is widely regarded as one of the most natural applications of quantum computers, since the cost of representing a correlated many-electron wavefunction grows exponentially with system size on classical hardware. In the current era of NISQ devices~\cite{Preskill2018}, where circuit depth is limited by decoherence and gate fidelity, fault-tolerant approaches such as quantum phase estimation remain out of reach. This has motivated the development of hybrid quantum-classical schemes. From these, the Variational Quantum Eigensolver (VQE)~\cite{Peruzzo2014,Tilly2022} was born as a reference algorithm for approximating molecular ground-state energies. VQE prepares a parametrized trial state on the quantum processor (QPU), estimates the expectation value of the molecular Hamiltonian, and updates the trial state parameters based on a classical optimiser, iterating until convergence. Within all the Variational Quantum Algorithms, VQE proposed a near-term solution based on the circuit requirements in the NISQ era~\cite{McClean2016,Cerezo2021}.

Firstly, the molecular Hamiltonian is expressed as a linear combination of Pauli strings, as in (\ref{eq:Hamiltonian_Pauli}) and (\ref{eq:Pauli_string}), under a fermion-to-qubit transformation such as the Bravyi-Kitaev~\cite{BravyiKitaev2002} mapping. The number of Pauli strings in a Hamiltonian scales as $\mathcal{O}(N^4)$ with the number of spin-orbitals $N$~\cite{Verteletskyi2020}. As the current hardware only supports projective measurements in the computational basis, the expectation value of $\mathcal{H}$ cannot be directly obtained and requires term-by-term reconstruction. Two Pauli strings can be simultaneously measured only if they are compatible. Non-commuting terms, consequently, require independent state preparations. This is the main issue that VQE presents, therefore, measurement reduction is a prerequisite for practical VQE rather than an optimisation detail.

\begin{equation}
    \label{eq:Hamiltonian_Pauli}
    \mathcal{H} = \sum c_iP_i
\end{equation}
\begin{equation}
    \label{eq:Pauli_string}
    P_i = \bigotimes p_j\quad;\quad p_j\in\left\{I, X, Y, Z\right\}
\end{equation}

Where $I, X, Y, Z$ are the Pauli matrices expressed as in Eq.~(\ref{eq:Pauli_matrices}):

\begin{equation}
    \label{eq:Pauli_matrices}
    \begin{aligned}
    I &= \begin{pmatrix} 1 & 0 \\ 0 & 1 \end{pmatrix}, &
    X &= \begin{pmatrix} 0 & 1 \\ 1 & 0 \end{pmatrix}, \\
    Y &= \begin{pmatrix} 0 & -i \\ i & 0 \end{pmatrix}, &
    Z &= \begin{pmatrix} 1 & 0 \\ 0 & -1 \end{pmatrix}.
    \end{aligned}
\end{equation}

This optimisation problem can be tackled analysing the commutativity graph of the Pauli Strings where each node of the graph represents a string and an edge connecting two nodes, $P$ and $Q$, means the two operators are compatible. This commutativity graph can be constructed on two types of relations: Full commutativity (FC), as in~(\ref{eq:full_commutativity}), or Qubit-Wise commutativity (QWC), as in~(\ref{eq:qwc}).

\begin{equation}\label{eq:full_commutativity}
    [P, Q]_{FC} \Leftrightarrow [P, Q] = 0
\end{equation}

\begin{equation}\label{eq:qwc}
    [P, Q]_{QWC}\Leftrightarrow [p_i, q_i] = 0\quad \forall\quad i
\end{equation}

Because pairwise commuting Hermitian operators can be simultaneously diagonalised, each set of pairwise compatible Pauli strings is jointly measurable and corresponds to a clique in the compatibility graph. Finding the smallest number of simultaneously measurable groups becomes exactly the Minimum Clique Cover (MCC) of the commutativity graph. MCC is an NP-hard~\cite{Karp1972} problem and since the minimum clique cover of a graph equals the chromatic number of its complement, the problem is addressed with classical colouring heuristics on that complement, typically greedy schemes such as DSATUR~\cite{Brelaz1979}. The same clique-cover formulation underlies the measurement-reduction schemes based on unitary transformations of fully commuting groups~\cite{Yen2020,Gokhale2020} and the $k$-commutativity relations that interpolate between the qubit-wise and full notions~\cite{dalfavero}, and the resource analysis of Ref.~\cite{Gonthier2022} quantifies how far the resulting measurement counts remain from practical quantum advantage in chemistry. The greedy heuristics are entirely adequate for the small systems on which VQE is usually demonstrated. For a molecule such as $H_2$ the commutativity graph is small and a greedy colouring is optimal in practice. The situation changes as the molecule grows. The number of vertices scales as $\mathcal{O}(N^4)$, the optimal number of groups grows with it, and classical heuristics lose the capability of guaranteeing optimality while exact classical solvers become prohibitively expensive. This work targets this regime, in which a better-than-greedy partition has a measurable impact and native quantum optimisation becomes an appealing candidate.

Quantum annealing is a natural candidate for this task, since it is a device designed to find the ground state of an Ising Hamiltonian~\cite{Kadowaki1998,Johnson2011} and an extensive family of NP-hard problems admits a Quadratic Unconstrained Binary Optimisation (QUBO) encoding~\cite{Lucas2014}. The colouring QUBO commonly used for this problem assigns one binary variable to each pair formed by a Pauli string and a candidate group, so it requires $M\cdot K$ logical variables for M Pauli strings and K candidate groups. This Ising formulation of the Pauli grouping problem has already been solved on fully connected CMOS Ising machines~\cite{kurita2022}, whose all-to-all coupling avoids any mapping step. A physical quantum annealer, in contrast, has a sparse fixed topology, so every logical variable must be embedded ~\cite{Choi2008, Cai2014} into a chain of physical qubits on a graph such as the Zephyr topology of the D-Wave Advantage2 processor ~\cite{Boothby2020}. That mapping step has been carried out on physical D-Wave processors~\cite{Jattana2024}. In previous works the same colouring QUBO was embedded in a triple-hybrid pipeline where the annealer supplies the measurement groups for a VQE run on gate-based hardware. The instances solved there carry at most a couple of hundred logical variables, built from model Hamiltonians, a truncated LiH term list and H$_2$ under both commutativity relations, and the fraction of valid samples is reported to decay with the variable count until two of the instances return no valid sample at all. The embedding that produces this decay is not characterised there. The interaction density of the QUBO governs not only its variable count but the size of the embedding, and whether the commutativity graphs of molecular Hamiltonians can be embedded at all on current annealers, and up to which molecular size, has not been systematically assessed. Even at the clique lower bound $\omega=13$, the smallest admissible colour budget, the LiH instance under full commutativity already carries $8190$ logical variables and $1.0\times10^{6}$ couplers, provably beyond what can be minor-embedded in the Zephyr topology; the resulting threshold for direct QPU sampling, which lies between H$_2$ and LiH, is the central limitation reported in this work and is analysed in detail in Sec.~\ref{sec:Embedding}.

In this work we address that gap. We adopt the standard colouring QUBO for the Minimum Vertex Colouring for the complement of the commutativity graph, also known as conflict graph. For a graded series of molecular systems of increasing complexity, and for both the qubit-wise and the full commutativity relations, we quantify the growth of the number of logical variables and the interaction density of the QUBO, and we characterise the embedding overhead, in physical qubits and chain length, onto the Zephyr topology of the Advantage2 architecture.This is the step left open by Jattana in Ref.~\cite{Jattana2024}, which samples the same encoding ona physical processor and observes the decay of valid samples with the variable count, but attributes the scale limitation of its pipeline to the gate-based device and never examines the embedding itself. The degree-counting bounds of Sec.~\ref{sec:bounds} settle non-embeddability without running any embedding heyristic, so the threshold reported below is a certificate on the encoding and the hardware graph rather than the outcome of a failed search. We then run the grouping problem on the annealer, both on the QPU through minor embedding and through hybrid quantum-classical solvers, and compare it against classical baselines, namely the greedy heuristics currently used in the VQE literature. Using time-to-solution and solution-quality metrics, this yields two distinct limits. The first is structural, the molecular size beyond which the problem can no longer be embedded in the processor topology. The second is practical, the size range over which annealing is worth using at a reasonable cost. The two do not coincide, and the order in which they are reached is the main practical result of this work, because the practical limit binds first. On H$_2$ under full commutativity, the only instance benchmarked on the QPU, sampling is already about twenty times more expensive in wall clock than the cheapest classical colouring for a partition that the same colouring finds exactly. Where a hybrid solver does return smaller covers than the best classical heuristic, it does so at two to three orders of magnitude more computation, and that advantage has already reversed on the largest instance it can process. The structural threshold therefore does not describe where a quantum annealer stops being useful for this task. It measures how far current hardware is from the size at which the question would become interesting.

%

\section{QUBO Formulation}\label{sec:QUBO}

Minimum Clique Cover on the commutativity graph $G$ and Minimum Vertex Colouring on its complement $\tilde{G}$ encode one and the same task, the search for the fewest simultaneously measurable groups, and they reach the same optimum, $\mathrm{MCC}(G)=\chi(\tilde{G})$, where $\chi$ denotes the chromatic number~\cite{Verteletskyi2020}. A clique of $G$ is an independent set of $\tilde{G}$, so a cover of $V$ by cliques of $G$ is a partition of $V$ into independent sets of $\tilde{G}$, that is, a proper colouring. The two formulations are, therefore, the same QUBO. Requiring a group to be a clique of $G$ penalises exactly the grouped pairs of strings that are not adjacent in $G$, and those are precisely the edges of $\tilde{G}$. Because both yield the same energy landscape and the same optimum, we develop only the colouring form, which in addition admits a leaner objective handling, and we do not detail the clique-cover model separately.

We work on $\tilde{G}=(V,\tilde{E})$, where $V$ is the set of $M$ Pauli strings and an edge of $\tilde{G}$ joins every two strings that are not compatible. We fix a number of candidate colours $K$, taken as an upper bound, and for each string $p\in V$ and colour $i\in\{1,\dots,K\}$ we introduce the binary variable defined in (\ref{eq:binary_mvc}), so that the model carries $M\cdot K$ logical variables.

\begin{equation}
    \label{eq:binary_mvc}
    x_{p,i} = \begin{cases}
        1 & \text{if string } p \text{ is coloured with colour } i\\
        0 & \text{otherwise}
    \end{cases}
\end{equation}

A configuration is a valid colouring when every string takes exactly one colour and no edge of $\tilde{G}$ joins two strings of the same colour, and each condition is imposed by a quadratic penalty~\cite{Lucas2014}. The first is the one-hot assignment penalty of (\ref{eq:one_hot_mvc}), which vanishes only when each string is given a single colour and grows quadratically with any string left uncoloured or assigned several colours at once.

\begin{equation}
\label{eq:one_hot_mvc}
\mathcal{H}_{\text{onehot}} = \alpha\sum_{p\in V}\left(1-\sum_{i=1}^{K} x_{p,i}\right)^{2}.
\end{equation}

The second enforces a proper colouring by charging an energy cost to every edge of $\tilde{G}$ whose two endpoints share a colour, as in (\ref{eq:coloring_mvc}), and it vanishes exactly when adjacent strings always receive different colours.

\begin{equation}
    \label{eq:coloring_mvc}
    \mathcal{H}_{\text{colouring}}=\beta\sum_{i=1}^{K}\ \sum_{\{p,q\}\in\tilde{E}} x_{p,i}\,x_{q,i}.
\end{equation}

The full model is the sum of the two penalties, (\ref{eq:QUBO_mvc}).

\begin{equation}\label{eq:QUBO_mvc}
    \mathcal{H}_{\text{MVC}} = \mathcal{H}_{\text{onehot}} + \mathcal{H}_{\text{colouring}}.
\end{equation}

Both weights are positive, $\alpha,\beta>0$, which is enough for the zero-energy criterion to be exact, since $\mathcal{H}_{\text{MVC}}=0$ demands that both penalties vanish at once. In practice we set $\alpha>\beta\,\Delta(\tilde{G})$, with $\Delta(\tilde{G})$ the maximum degree of the complement, so that leaving a string uncoloured is never cheaper than the conflicts it would avoid and the low-energy samples returned by the annealer keep the one-hot structure even when the chosen $K$ admits no proper colouring.

Equation (\ref{eq:QUBO_mvc}) is a decision model at fixed $K$: its ground-state energy is zero if and only if $\tilde{G}$ admits a proper colouring with $K$ colours, equivalently if and only if the $M$ Pauli strings can be partitioned into $K$ simultaneously measurable groups. The minimisation is recovered by an outer loop over $K$. Starting from the greedy upper bound, we lower $K$ by one and re-solve while a zero-energy configuration is still found. For an exact solver the smallest value of $K$ would be $\chi(\tilde{G})$, and therefore the minimum clique cover of $G$. With a heuristic sampler it is only an upper bound on it, since the absence of a zero-energy sample at $K-1$ does not prove that no proper colouring exists. We denote it $K_{\text{found}}$ throughout. This keeps every submitted QUBO at $M\cdot K$ variables and two penalty terms, without the auxiliary variables that an explicit count of the used colours would require, an advantage that matters here because the size and interaction density of the QUBO govern its embedding onto the annealer.

\subsection{Molecular instances}\label{subsec:instances}

The model of Equation~(\ref{eq:QUBO_mvc}) is evaluated on a set of six molecules of increasing size, used throughout Sections~\ref{sec:Embedding} and~\ref{sec:Benchmarks}. All systems are treated in the STO-3G basis. H$_2$ ($d = 0.7414~\text{\AA}$) and LiH ($d =1.5949~\text{\AA}$) are taken at their experimental equilibrium geometries. For BeH$_2$ (collinear, $d = 1.4~\text{\AA}$), H$_2$O ($r = 0.75~\text{\AA}$, $\theta = 107.6^\circ$), NH$_3$ (pyramidal, $r = 1.0~\text{\AA}$, $\theta_{\text{HNH}} = 107.0^\circ$) and N$_2$ ($d = 1.1~\text{\AA}$) we adopt the geometries of Ref.~\cite{Verteletskyi2020}, since the number of non-vanishing Pauli strings is sensitive to molecular symmetry and these values make our term counts directly comparable with those reported there. The symmetric coordinates of H$_2$O and NH$_3$ are computed from the bond length and angle at full floating-point precision rather than tabulated in rounded Cartesian form, because rounding is enough to break the C$_{2v}$ and C$_{3v}$ symmetry of the Hartree-Fock solution and to change the term count.

The Hartree-Fock solution and the second-quantised Hamiltonian are obtained with the differentiable Hartree-Fock implementation of PennyLane~\cite{pennylane} in the full orbital space with no frozen core and no active-space reduction. The Hamiltonian is mapped to qubits with the Jordan-Wigner transformation~\cite{JordanWigner1928}, giving 4, 12, 14, 14, 16 and 20 qubits respectively. Pauli coefficients whose magnitude falls below $10^{-6}$ are discarded. The identity string is not included in the term count $M$, as it requires no measurement and in the conflict graph $\tilde{G}$ it is an isolated vertex. Under this convention the six instances contain $ M = 14, 630, 665, 1085, 3608, 2950$ Pauli strings.

We note in passing that the same molecules, treated with symmetry-adapted orbitals, produce fewer Pauli strings at identical Hartree-Fock energy. The choice of orbital representation acts on the Hamiltonian rather than on the grouping problem itself, and combining the two would confound the scaling this work sets out to measure, keeping the canonical full-space Hamiltonians also leaves the instance sizes directly recognisable against those used elsewhere in the measurement grouping literature. Since the colouring QUBO carries $M\cdot K$ variables, however, a reduction of that magnitude in $M$ acts directly on the quantity that limits embeddability. Table~\ref{tab:conflict} summarises the resulting conflict graphs.

%

\section{Embedding} \label{sec:Embedding}
To run the model of (\ref{eq:QUBO_mvc}) on a physical annealer, its interaction graph must be minor-embedded into the hardware graph: every logical variable is mapped to a connected chain of physical qubits held together by a strong ferromagnetic coupling, and every quadratic term must be realised by at least one physical coupler joining the two corresponding chains~\cite{Choi2008,Cai2014}. Two hardware generations are considered here. The Zephyr graph $Z_m$ of the Advantage2 generation has $16m(2m+1)$ qubits of degree at most $20$~\cite{BoothbyZephyr2021}, so the ideal $Z_{12}$ behind the current processors has $4\,800$ qubits and $45\,864$ couplers; the Pegasus graph $P_m$ of the previous Advantage generation has qubits of degree at most $15$~\cite{Boothby2020}, and its fabricated form, obtained by discarding the $8(m-1)$ incomplete boundary qubits of the $24m(m-1)$ of the full lattice, carries $24m(m-1)-8(m-1)$ qubits, that is $5\,640$ qubits and $40\,484$ couplers for $P_{16}$. The processor targeted in this work is the \texttt{Advantage2\_system1} solver, whose working graph, retrieved through the Leap API in July 2026, is a defect-laden subgraph of $Z_{12}$ with $4\,577$ qubits and $41\,515$ couplers, that is, $95\%$ of the ideal qubits and $91\%$ of the ideal couplers~\cite{DWaveAdv2}. Throughout this section, embeddings on the ideal graphs isolate the topological cost of the QUBO from fabrication defects, and the working graph quantifies the additional cost of the real chip.

\subsection{The source graph is a Cartesian product}\label{sec:cartesian}

The graph the annealer has to host is not the conflict graph $\tilde{G}$ but the interaction graph $S$ of the QUBO, and $S$ has an exact product structure. Expanding the one-hot penalty of (\ref{eq:one_hot_mvc}) couples, for every string $p$, all pairs $x_{p,i},x_{p,j}$ with $i\neq j$, one complete graph $K_K$ per string; the colouring penalty of Eq.~(\ref{eq:coloring_mvc}) couples $x_{p,i}$ and $x_{q,i}$ for every edge $\{p,q\}\in\tilde{E}$ and every colour $i$, that is, $K$ disjoint copies of $\tilde{G}$. These are precisely the edges of the Cartesian product

\begin{equation}\label{eq:source_graph}
    S(\tilde{G},K) \;=\; \tilde{G}\,\square\,K_K,
\end{equation}

so its size and degree sequence are known in closed form,

\begin{equation}\label{eq:source_counts}
    |V(S)| = M\,K, \qquad |E(S)| = \frac{M\,K(K-1)}{2} + K\,|\tilde{E}|,
\end{equation}

\begin{equation}\label{eq:source_degree}
    \deg_S(x_{p,i}) = (K-1) + \deg_{\tilde{G}}(p).
\end{equation}

Here $K$ denotes a generic colour budget; from Subsec.~\ref{sec:conflictgraphs} onwards it is evaluated at $\omega$.

Equations~(\ref{eq:source_counts}) and~(\ref{eq:source_degree}) quantify what the QUBO encoding itself costs before any hardware enters the picture. The annealer never sees the $M$-node graph that a classical heuristic colours directly; it sees a graph $K$ times larger, in which between $2\%$ and $38\%$ of the couplers (the one-hot cliques, first term of Eq.~(\ref{eq:source_counts})) exist solely to enforce feasibility and carry no information about $\tilde{G}$. The maximum degree $\Delta(S)=K-1+\Delta(\tilde{G})$ exceeds the hardware degree $20$ for every instance studied beyond H$_2$ (Table~\ref{tab:source}), so chains are unavoidable there. They are in fact unavoidable much earlier: the clique number of the ideal $Z_{12}$ graph is $4$, established by exhaustive enumeration of its maximal cliques, of which $8\,464$ are copies of $K_4$ and none is larger, in agreement with the four-qubit cliques described in Ref.~\cite{BoothbyZephyr2021}; since every working graph is a subgraph of the ideal fabric, its clique number is at most $4$ as well. Every one-hot clique with $K\geq 5$ therefore forces at least one chain of length two or more, independently of how sparse $\tilde{G}$ is. Figure~\ref{fig:h2_embedding}(a) makes this structure visible for the smallest instance, H$_2$ under QWC: five stacked copies of $\tilde{G}$, one per candidate colour, pierced by fourteen one-hot skewers, one per Pauli string.

\begin{remark}\label{rem:peeling}
The source graph at $K-1$ colours is the subgraph of $S(\tilde{G},K)$ induced by deleting one colour layer. Any embedding of $S(\tilde{G},K)$ therefore restricts, by simply discarding the chains of that layer, to a valid embedding of $S(\tilde{G},K-1)$: the outer loop over $K$ of Sec.~\ref{sec:QUBO} needs a single embedding computed once at the greedy upper bound, not one search per value of $K$.
\end{remark}

\subsection{Conflict graphs}\label{sec:conflictgraphs}

Table~\ref{tab:conflict} lists the conflict graphs $\tilde{G}$ of the six instances of Subsec.~\ref{subsec:instances}, for both commutativity relations, together with the greedy clique lower bound $\omega\leq\chi(\tilde{G})$, a structural property of the graph that will certify optimality where it meets the colour budget. The colouring side of the problem, heuristic and quantum, is the subject of Sec.~\ref{sec:Benchmarks} and enters this section only through the colour budget, which fixes the size of the QUBO to be embedded. We evaluate that size at $\omega$ throughout, so that every impossibility reported below survives any improvement in the colouring heuristics. FC needs $2.5$ to $58$ times fewer groups than QWC on the same molecule, and, through the quadratic dependence of Eq.~(\ref{eq:source_counts}) on $K$, this single factor separates the two relations by more than an order of magnitude in QUBO couplers.

\begin{table*}[t]
\caption{\label{tab:conflict}Conflict graphs $\tilde{G}$ of the benchmark molecules of Subsec.~\ref{subsec:instances} (STO-3G, Jordan-Wigner, full orbital space). $n_q$ is the number of qubits and $M$ the number of Pauli strings.
 $|\tilde{E}|$ and $\Delta(\tilde{G})$ the edge count and maximum degree of the conflict graph, and $\omega$ a greedy clique lower bound on the chromatic number $\chi(\tilde{G})$.}
\begin{ruledtabular}
\begin{tabular}{lllrrrr}
Molecule & $n_q$ & Rel. & $M$ & $|\tilde{E}|$ & $\Delta(\tilde{G})$ & $\omega$ \\
\hline
H$_2$ & 4 & FC & 14 & 16 & 4 & 2 \\
H$_2$ & 4 & QWC & 14 & 46 & 13 & 5 \\
LiH & 12 & FC & 630 & 76\,272 & 264 & 13 \\
LiH & 12 & QWC & 630 & 153\,192 & 593 & 138 \\
BeH$_2$ & 14 & FC & 665 & 78\,408 & 268 & 12 \\
BeH$_2$ & 14 & QWC & 665 & 162\,532 & 609 & 202 \\
H$_2$O & 14 & FC & 1085 & 214\,056 & 428 & 16 \\
H$_2$O & 14 & QWC & 1085 & 474\,762 & 1029 & 308 \\
NH$_3$ & 16 & FC & 3608 & 2\,274\,848 & 1488 & 26 \\
NH$_3$ & 16 & QWC & 3608 & 5\,728\,956 & 3529 & 1128 \\
N$_2$ & 20 & FC & 2950 & 1\,272\,112 & 936 & 19 \\
N$_2$ & 20 & QWC & 2950 & 3\,642\,982 & 2813 & 1107 \\

\end{tabular}
\end{ruledtabular}
\end{table*}

\subsection{Degree-counting lower bounds}\label{sec:bounds}

Whether an instance fits on a processor can be settled, negatively, without running any embedding heuristic. In a hardware graph of maximum degree $D$, a chain of $L$ qubits has at most $DL-2(L-1)=(D-2)L+2$ couplers leaving the chain, so a logical variable of degree $d_v$ needs a chain of length

\begin{equation}\label{eq:chain_bound}
    L_v \;\geq\; \max\!\left(1,\;\Big\lceil \tfrac{d_v-2}{D-2} \Big\rceil\right),
\end{equation}

and, because chains of different variables are vertex-disjoint, the per-variable bounds add up to a bound on the physical qubits of any valid embedding,

\begin{equation}\label{eq:qubit_bound}
    Q_{\min} \;=\; \sum_{v\in V(S)} \max\!\left(1,\;\Big\lceil \tfrac{d_v-2}{D-2} \Big\rceil\right) \;\leq\; Q.
\end{equation}

Couplers are counted the same way: every source edge occupies at least one inter-chain coupler, every chain consumes $L_v-1$ internal couplers, and all of these are distinct, so the hardware must supply at least $|E(S)|+Q-|V(S)|$ couplers. These bounds use only the degree sequence of Eq.~(\ref{eq:source_degree}), ignore the geometry of the topology entirely, and are therefore loose; the true requirements can only be larger.

Table~\ref{tab:source} applies them to every instance for the Zephyr degree $D=20$ and the Pegasus degree $D=15$. The verdict is unambiguous: every instance beyond H$_2$ exceeds the $4\,800$ qubits of the full ideal $Z_{12}$ by a factor between $23$ (BeH$_2$, FC) and $2.0\times10^{5}$ (NH$_3$, QWC), and the $4\,577$ of the working chip by a factor between $25$ and $2.1\times10^{5}$.
Non-embeddability on current hardware is thus a certificate, not a heuristic failure, and no embedding search is needed to rule these instances out. Feeding $Q_{\min}$ and the coupler bound back into the qubit and coupler counts of the ideal families yields the minimal generation index $m^{*}$ at which each instance could possibly fit: the smallest infeasible instance, BeH$_2$ under FC, needs at least $Z_{59}$, that is $\approx1.1\times10^{5}$ qubits or the fabric of $25$ Advantage2 processors, and the requirement grows to $Z_{5518}$ for NH$_3$ under QWC. The Pegasus columns are uniformly $35$--$40\%$ more expensive, quantifying, at the level of provable bounds, the connectivity gain of the newer topology.

The H$_2$ QWC instance measures how loose: its bounds are met by $Z_2$, yet no embedding is found there in forty seeded attempts, and the smallest Zephyr graph that hosts it is $Z_3$, twice the qubits. Part of the gap is that $D=20$ is the asymptotic degree of the family: $Z_2$ has maximum degree $19$, with $64$ of its $160$ qubits at degree $10$ or $11$, so the bound is evaluated at a degree the graph never attains. The $m^{*}$ values are therefore lower bounds on the generation index, and the verdict they support is the negative one.

The same bounds settle whether the obstruction lies in the problem or in the way it is written as a QUBO. Applying Eq.~(\ref{eq:qubit_bound}) to the conflict graph itself rather than to the source graph gives a floor on the qubits of any encoding in which the $M$ strings occupy disjoint connected sets of physical qubits and every conflict of $\tilde{E}$ is realised by at least one coupler between the corresponding sets, a condition met by one-hot and domain-wall~\cite{Chancellor2019} encodings alike, and violated by encodings that make a conflict non-quadratic and therefore pay for ancillas. Since $\sum_{p}\lceil(\deg_{\tilde{G}}(p)-2)/(D-2)\rceil\geq(2|\tilde{E}|-2M)/(D-2)$, that floor is already $8\,405$ qubits for LiH under FC and $8\,639$ for BeH$_2$, both above the $4\,800$ of the ideal $Z_{12}$ and the $4\,577$ of the working chip, and it reaches $6.4\times10^{5}$ for NH$_3$ under QWC. Beyond H$_2$ the conflict graphs are thus too dense for the topology on their own, with mean degrees between $236$ and $3\,176$ against a hardware degree of $20$, and no encoding in this class removes that. What the one-hot encoding adds on top is a factor of order $K$: the ratio of $Q_{\min}$ in Table~\ref{tab:source} to this floor is $13$ to $27$ under FC and $180$ to $1.6\times10^{3}$ under QWC. The negative verdict is therefore twice determined, once by the density of $\tilde{G}$ and once by the cost of the colour variables, and only the second is open to reformulation.

\begin{table*}[t]
\caption{\label{tab:source}Source graphs $S=\tilde{G}\,\square\,K_K$ of the colouring QUBO and degree-counting lower bounds on any minor embedding, for the Zephyr ($D=20$) and Pegasus ($D=15$) topologies. The source graph is evaluated at $\omega$, the clique lower bound reported in Table~\ref{tab:conflict} and reused verbatim here: since $\omega\leq\chi(\tilde{G})$, no admissible colour budget is smaller, so this is the smallest QUBO the instance could ever require and the bounds below are the most favourable ones it admits. The budget actually submitted, a greedy upper bound on $\chi(\tilde{G})$, is larger, and would only make every entry grow. $L_{\min}$ bounds the longest chain via Eq.~(\ref{eq:chain_bound}) applied to $\Delta(S)$, $Q_{\min}$ is the qubit bound of Eq.~(\ref{eq:qubit_bound}), and $m^{*}$ is the smallest generation index whose ideal qubit and coupler counts satisfy both bounds. The full ideal $Z_{12}$ of Advantage2 has $4\,800$ qubits; only H$_2$ fits.}
\begin{ruledtabular}
\begin{tabular}{ll|rrrr|rrr|rrr}
 & & & & & & \multicolumn{3}{c|}{Zephyr ($D=20$)} & \multicolumn{3}{c}{Pegasus ($D=15$)} \\
Molecule & Rel. & $\omega$ & $|V(S)|$ & $|E(S)|$ & $\Delta(S)$ & $L_{\min}$ & $Q_{\min}$ & $m^{*}$ & $L_{\min}$ & $Q_{\min}$ & $m^{*}$ \\
\hline
H$_2$ & FC & 2 & 28 & 46 & 5 & 1 & 28 & 1 & 1 & 28 & 2 \\
H$_2$ & QWC & 5 & 70 & 370 & 17 & 1 & 70 & 2 & 2 & 90 & 3 \\
LiH & FC & 13 & 8\,190 & 1\,040\,676 & 276 & 16 & 119\,782 & 61 & 22 & 164\,814 & 84 \\
LiH & QWC & 138 & 86\,940 & 27\,095\,886 & 730 & 41 & 3\,044\,556 & 309 & 56 & 4\,194\,096 & 419 \\
BeH$_2$ & FC & 12 & 7\,980 & 984\,786 & 279 & 16 & 112\,164 & 59 & 22 & 155\,820 & 82 \\
BeH$_2$ & QWC & 202 & 134\,330 & 46\,331\,629 & 810 & 45 & 5\,195\,844 & 403 & 63 & 7\,166\,354 & 547 \\
H$_2$O & FC & 16 & 17\,360 & 3\,555\,096 & 443 & 25 & 404\,176 & 113 & 34 & 553\,552 & 153 \\
H$_2$O & QWC & 308 & 334\,180 & $1.98\times10^{8}$ & 1\,336 & 75 & 22\,062\,040 & 831 & 103 & 30\,475\,368 & 1128 \\
NH$_3$ & FC & 26 & 93\,808 & 60\,318\,648 & 1\,513 & 84 & 6\,731\,608 & 459 & 117 & 9\,310\,288 & 624 \\
NH$_3$ & QWC & 1128 & 4\,069\,824 & $8.76\times10^{9}$ & 4\,656 & 259 & $9.74\times10^{8}$ & 5518 & 358 & $1.35\times10^{9}$ & 7496 \\
N$_2$ & FC & 19 & 56\,050 & 24\,674\,578 & 954 & 53 & 2\,753\,518 & 294 & 74 & 3\,822\,040 & 400 \\
N$_2$ & QWC & 1107 & 3\,265\,650 & $5.84\times10^{9}$ & 3\,919 & 218 & $6.50\times10^{8}$ & 4507 & 302 & $8.99\times10^{8}$ & 6122 \\
\end{tabular}
\end{ruledtabular}
\end{table*}

\subsection{Embedding the feasible instance}\label{sec:h2embed}
 
The all-to-all capacity of a topology is the size of the largest complete graph it can host, and structured clique embeddings give its operational value: the \texttt{busclique} generator of \texttt{minorminer}~\cite{Cai2014,minorminerSW} places $K_{184}$ on the ideal $Z_{12}$ with uniform chains of $13$ qubits ($2\,392$ qubits in total), against $K_{180}$ with chains of up to $17$ on the ideal $P_{16}$: one generation buys a quarter shorter chains at comparable clique size, consistent with the cross-generation measurements of Ref.~\cite{Pelofske2025}. On the working graphs the picture inverts, because fabrication yield dominates: the \texttt{Advantage2\_system1} working graph retains $95\%$ of its qubits, yet its defects cut the clique ceiling to $K_{106}$ (chains of at most $13$), while an Advantage-generation solver still online in Leap (\texttt{Advantage\_system6}, $5\,612$ working qubits, $99.5\%$ of the $P_{16}$ fabric, $40\,088$ couplers) preserves $K_{175}$ with chains of $17$. 
On today's processors the older chip therefore hosts the larger complete graph, and the connectivity advantage of Zephyr materialises in chain length rather than clique size. Any QUBO whose variable count exceeds this clique ceiling and whose density approaches one is out of reach; the colouring QUBOs live far from that regime, with densities of $10^{-3}$ to $10^{-1}$, so sparsity-exploiting embedding can and does beat the clique route where it applies.
 
\begin{figure*}[t]
    \centering
    \subfloat[Source graph $S(\tilde{G},5)$ of H$_2$, QWC\label{fig:h2emb_a}]{%
        \includegraphics[width=0.32\textwidth]{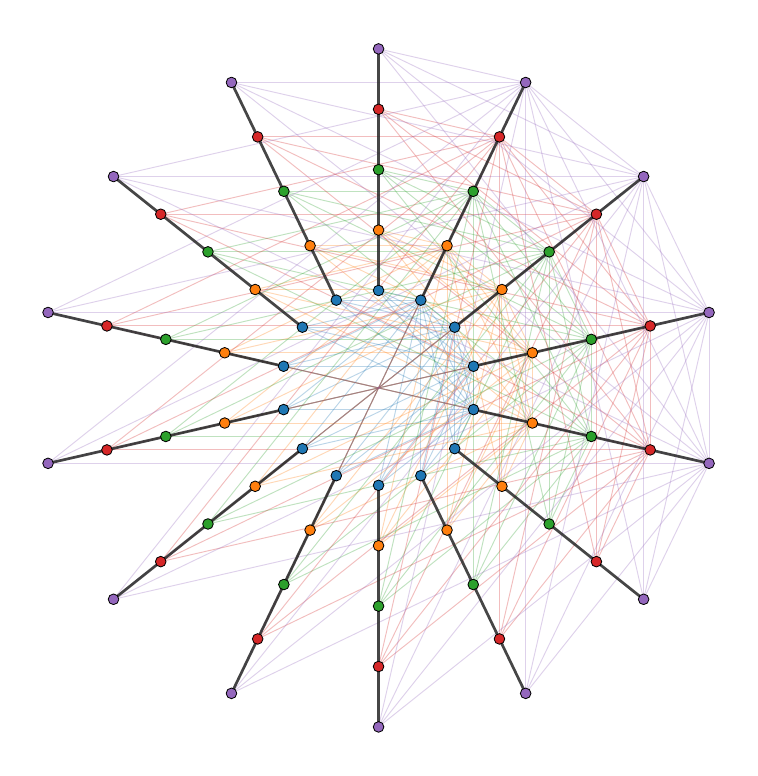}%
    }\hfill
    \subfloat[Minor embedding into $Z_3\subset Z_{12}$\label{fig:h2emb_b}]{%
        \includegraphics[width=0.32\textwidth]{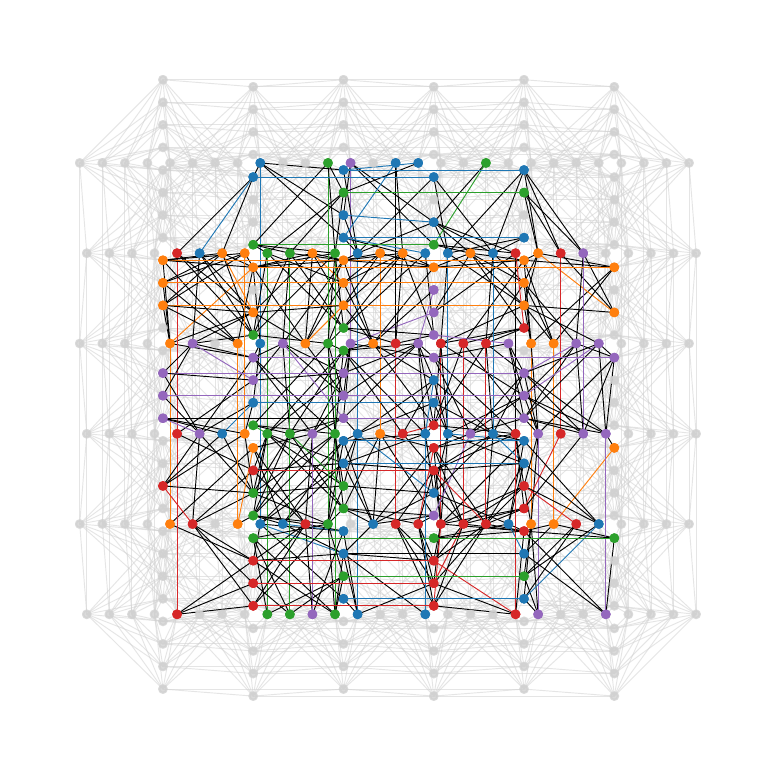}%
    }\hfill
    \subfloat[H$_2$ FC, native embedding into $Z_1$\label{fig:h2emb_c}]{%
        \includegraphics[width=0.32\textwidth]{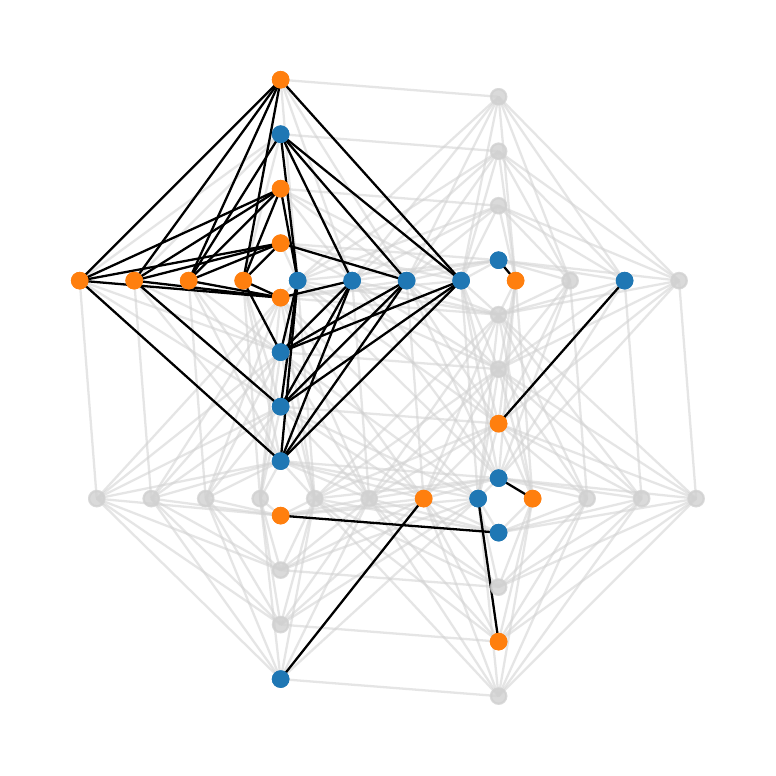}%
    }
    \caption{Embedding the feasible instances, drawn with the Ocean SDK tooling~\cite{dwave_ocean_sdk,minorminerSW}. (a) The source graph of H$_2$ under QWC as the Cartesian product of Eq.~(\ref{eq:source_graph}): five concentric copies of the conflict graph $\tilde{G}$, one per candidate colour (chords, one colour per layer), pierced by one one-hot clique $K_5$ per Pauli string, whose five nodes are collinear so that its ten edges overlap on a single radial spoke. (b) Minor embedding of (a) into the ideal $Z_3$ graph ($336$ qubits), an induced subgraph of the ideal $Z_{12}$ fabric, so the embedding shown carries over verbatim to the processor-scale graph; found by \texttt{minorminer} in a dedicated $30$-seed sweep on $Z_3$ (seed $17$ shown): $178$ physical qubits, longest chain $5$, mean chain $2.54$, not directly comparable with the $Z_{12}$ medians of Table~\ref{tab:h2embed}, which target the full fabric; each chain inherits the colour of its layer, unused qubits and couplers in grey. (c) The FC instance ($\omega=2$, $28$ variables) embedded in the smallest Zephyr graph $Z_1$ ($48$ qubits, likewise an induced subgraph of $Z_{12}$): every chain is a single physical qubit, that is, the QUBO is a subgraph of the hardware.}
    \label{fig:h2_embedding}
\end{figure*}

Table~\ref{tab:h2embed} reports the embeddings of the two feasible instances, obtained with \texttt{minorminer} (ten seeds on the ideal targets, twenty on the working graphs, sixty-second budget, every embedding re-verified independently for chain connectivity, chain disjointness and edge coverage) and with \texttt{busclique} treating the QUBO as a complete graph on $|V(S)|$ nodes. Three facts stand out. First, H$_2$ under FC ($\omega=2$, $28$ variables) embeds natively: all chains have length one, the QUBO is a subgraph of Zephyr, and $28$ physical qubits suffice, a configuration that even fits the smallest member $Z_1$ of the family ($48$ qubits, Fig.~\ref{fig:h2_embedding}(c)) and, more importantly, survives fabrication defects, since the same chain-free embedding is found on the working chip in all twenty seeds. Second, H$_2$ under QWC ($\omega=5$, $70$ variables) cannot embed natively on any Zephyr graph, since $\omega=5$ exceeds the clique number $4$ of the topology; \texttt{minorminer} needs a median of $206$ qubits and a median longest chain of $5$ (best seed: $4$) on the ideal $Z_{12}$, rising mildly to $212$ qubits and chains of $6$ on the working chip, while on Pegasus it needs some $250$ qubits with a best-seed longest chain of $6$, on the ideal fabric and on the working solver alike: at $99.5\%$ yield the defects are not visible above the seed-to-seed scatter, whose interquartile range spans around twenty qubits. Figure~\ref{fig:h2_embedding}(b) shows one such chained embedding. Third, at these densities ($0.12$--$0.15$) the generic heuristic beats the structured clique construction by a factor $2$ to $4$ in qubits ($206$ vs.\ $420$ on the ideal $Z_{12}$; $28$ vs.\ $98$ on the working Advantage2), the reverse of the dense regime \texttt{busclique} is designed for. The minimal-hardware sweep sharpens the geometric picture: the ideal graphs are nested as induced subgraphs, $Z_1\subset Z_2\subset\cdots\subset Z_{12}$ (verified computationally via the Zephyr coordinates), so embeddability is monotone in the generation index and the smallest feasible generation is well defined; QWC first embeds on $Z_3$ ($336$ qubits) and fails in forty seeded attempts on $Z_2$ ($160$ qubits), while on Pegasus it first embeds on $P_5$ and fails on $P_4$ even though $P_4$ carries $264$ qubits, more than the $206$-qubit embedding found on $Z_{12}$: qubit count alone does not decide feasibility, geometry does.
 
\begin{table*}[t]
\caption{\label{tab:h2embed}Verified embeddings of the two feasible instances on ideal and working targets. \texttt{minorminer} entries are medians (interquartile range in brackets for qubits) over ten seeds on ideal targets and twenty on working graphs, all from a single environment~\cite{minorminerSW}; \texttt{busclique} is deterministic. $\ell_{\max}$ and $\bar{\ell}$ are the maximum and mean chain length. Working graphs retrieved through the Leap API in July 2026: \texttt{Advantage2\_system1} ($4\,577$ qubits, $41\,515$ couplers) and \texttt{Advantage\_system6} ($5\,612$ qubits, $40\,088$ couplers).}
\begin{ruledtabular}
\begin{tabular}{ll|rrrr|rr}
 & & \multicolumn{4}{c|}{\texttt{minorminer}} & \multicolumn{2}{c}{\texttt{busclique}} \\
Instance & Target & Qubits & $\ell_{\max}$ & $\bar{\ell}$ & $t$ (s) & Qubits & $\ell_{\max}$ \\
\hline
H$_2$ FC ($|V(S)|=28$) & $Z_{12}$ ideal & 28 [28, 28] & 1 & 1.00 & 0.04 & 112 & 4 \\
H$_2$ FC ($|V(S)|=28$) & Advantage2 working & 28 [28, 28] & 1 & 1.00 & 0.04 & 98 & 4 \\
H$_2$ FC ($|V(S)|=28$) & $P_{16}$ ideal & 30 [30, 30] & 2 & 1.07 & 0.04 & 106 & 4 \\
H$_2$ FC ($|V(S)|=28$) & Advantage working & 30 [30, 32] & 2 & 1.07 & 0.05 & 106 & 4 \\
H$_2$ QWC ($|V(S)|=70$) & $Z_{12}$ ideal & 206 [205, 212] & 5 & 2.94 & 1.7 & 420 & 6 \\
H$_2$ QWC ($|V(S)|=70$) & Advantage2 working & 212 [208, 223] & 6 & 3.03 & 1.4 & 486 & 7 \\
H$_2$ QWC ($|V(S)|=70$) & $P_{16}$ ideal & 254 [245, 265] & 7 & 3.63 & 1.9 & 548 & 8 \\
H$_2$ QWC ($|V(S)|=70$) & Advantage working & 248 [243, 268] & 6 & 3.55 & 1.9 & 550 & 8 \\
\end{tabular}
\end{ruledtabular}
\end{table*}
 
The structural limit promised in Sec.~\ref{sec:Introduction} is therefore located with precision: of the twelve molecular instances, exactly two are embeddable on current annealing hardware, both belonging to H$_2$, and for both the clique bound of Table~\ref{tab:conflict} is already attained by the greedy colourings of Sec.~\ref{sec:Benchmarks} ($\omega=K_{\text{found}}=\chi$), so the optimum is certified before any sampling. The QPU experiments of Sec.~\ref{sec:Benchmarks} consequently measure sampling efficiency, the time and reliability with which the annealer reaches a valid $K$-colouring, rather than partition-quality gains; the instances where quality gains over greedy heuristics are conceivable, namely the FC instances whose greedy colourings in Table~\ref{table:results} stay well above $\omega$, sit two to five orders of magnitude beyond the embeddable regime.

%
\section{Optimisation Benchmarks} 
\label{sec:Benchmarks}
\subsection{Benchmark setup}
The benchmarks in this section use the six Hamiltonians defined in Section~\ref{subsec:instances} and the conflict graphs $\tilde{G}$ of Table~\ref{tab:conflict}, built under both commutativity relations. Every column of the tables below, classical and quantum, was produced in this work over the same graphs, so that differences between methods reflect the method and not the instance.

The colouring of $\tilde{G}$ is addressed in two ways. Classically, we use three greedy heuristics: Largest First (LF)~\cite{LF_algorithm}, Recursive Largest First (RLF)~\cite{RLF_algorithm}, and the vertex-merging scheme of Dutton and Brigham (DB)~\cite{DB_algorithm}. All three are run with a fixed tie-breaking rule, so they are deterministic and return a single value per instance. For the annealing approach, the one-hot QUBO of Eq.~(\ref{eq:QUBO_mvc}) is submitted to three backends via the D-Wave Ocean SDK~\cite{dwave_ocean_sdk}: simulated annealing (SA) on a classical processor through the \texttt{neal} package, direct sampling on the \texttt{Advantage2\_system1} QPU with minor embedding onto the Zephyr topology (QA), and the Leap Hybrid BQM solver (Hybrid). Only direct QPU sampling uses quantum hardware; \texttt{neal} is a classical algorithm, and the hybrid solver combines classical preprocessing with quantum sampling. Neither SA nor Hybrid is presented here as a QPU result.

Following the outer loop of Section~\ref{sec:QUBO}, $K$ is initialised at the upper bound returned by LF and decreased by one while a zero-energy sample is still returned, terminating at the first value for which no read succeeds or at a value that is trivially infeasible. The quantity reported is $K_{\text{found}}$, an upper bound on the minimum number of groups and never the chromatic number itself, as failing to reach zero energy at $K-1$ does not prove infeasibility.

\subsection{The commutativity graph of H$_2$}\label{sec:h2graph}
\begin{figure*}
    \centering
    \subfloat[Full commutativity\label{fig:h2_graph_fc}]{%
        \includegraphics[width=0.48\textwidth]{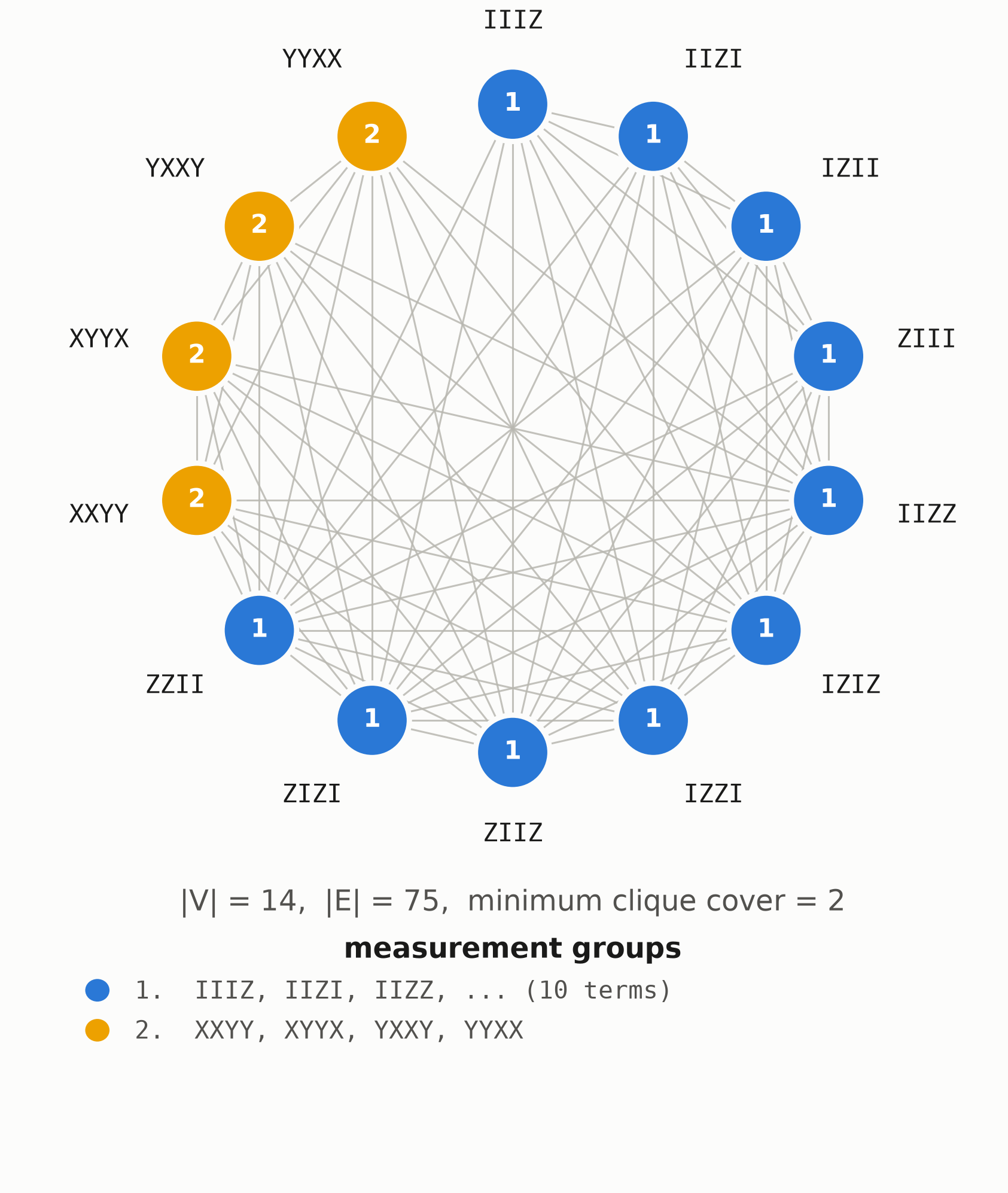}%
    }\hfill
    \subfloat[Qubit-wise commutativity\label{fig:h2_graph_qwc}]{%
        \includegraphics[width=0.48\textwidth]{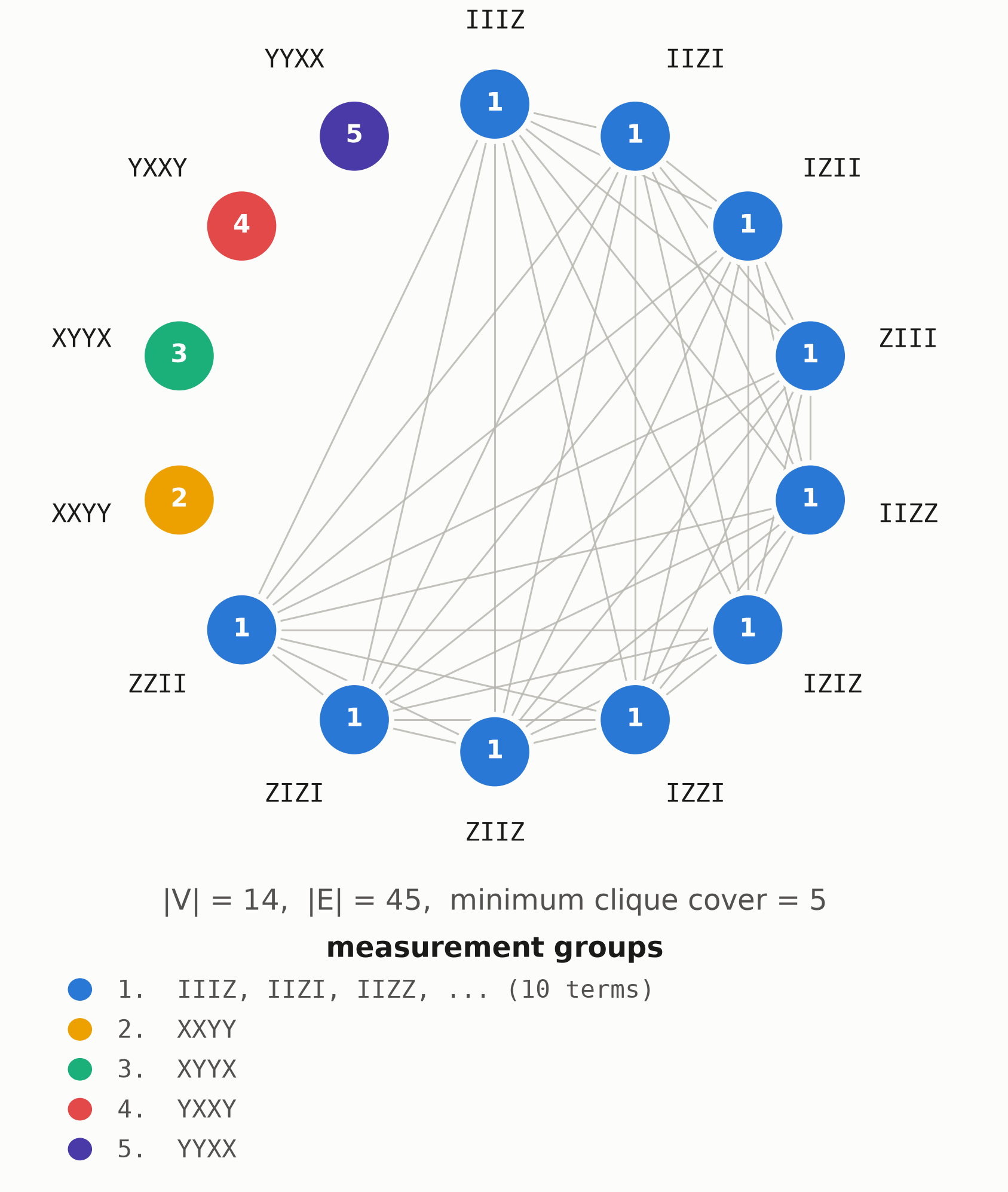}%
    }
    \caption{Minimum clique cover of the two commutativity graphs of the H$_2$/STO-3G
        Hamiltonian under the Jordan-Wigner transformation (14 Pauli strings, identity excluded). (a) Full commutativity, 75 edges, two groups. (b) Qubit-wise commutativity, 45 edges, five groups.}
    \label{fig:h2_graph}
\end{figure*}

Figure~\ref{fig:h2_graph} shows the two commutativity graphs of the H$_2$ Hamiltonian. Under full commutativity (FC), shown in Figure~\ref{fig:h2_graph_fc}, the fourteen Pauli strings are joined by 75 edges, and the exact minimum clique cover consists of two groups: the ten strings built from identity and Z operators, which commute pairwise, and the four strings $XXYY, XYYX, YXXY$, and $YYXX$, which commute with one another but not qubit-wise. Under qubit-wise commutativity (QWC), shown in Figure~\ref{fig:h2_graph_qwc}, thirty of those edges disappear, leaving 45, and the cover grows to five groups. In terms of the complement, the conflict graph on which the colouring QUBO acts has 16 edges under FC and 46 under QWC, as listed in Table~\ref{tab:conflict}.

This instance illustrates the two effects that drive the rest of the section. First, FC produces much denser commutativity graphs and therefore a sparser conflict graph, which yields fewer measurement groups and, since the model carries $M\cdot K$ variables, a smaller QUBO for the same Hamiltonian. Second, the gain is not uniform; it comes entirely from the Pauli strings with several non-identity, non-Z factors, which are precisely those that qubit-wise commutativity is forced to isolate. For H$_2$, the clique lower bound already meets the cover under both relations, with $\omega=\chi=2$ and $5$, respectively. This instance is therefore solved to certified optimality by every method considered below and provides no information about their relative quality.

\subsection{Solution Quality}

\begin{table}
    \centering
    \caption{Number of fully commuting measurement groups obtained for the six instances of Section~\ref{subsec:instances}. $n_q$ is the number of qubits needed for the VQE after the Jordan-Wigner transformation. $M$ is the number of Pauli strings. RLF, LF, and DB are the classical greedy colourings, while SA, QA, and Hybrid denote simulated annealing, direct sampling on the Advantage2 QPU, and the Leap Hybrid BQM solver, respectively. All entries are $K_{\text{found}}$, an upper bound on the minimum number of groups. A dash denotes an instance that the corresponding sampler could not solve or embed. All results are obtained after a single execution of the algorithm.}
    \resizebox{\columnwidth}{!}{%
    \begin{ruledtabular}
    \begin{tabular}{l c c | c c c | c c c}
        System & $n_q$ & $M$ & RLF & LF & DB  & SA & QA & Hybrid \\
        \hline
        H$_2$  & 4 & 14   & 2  & 2   & 2  & 2  &  2 & 2\\
        LiH    &12 & 630  & 26 & 35  & 24 & -- & -- & 20\\
        BeH$_2$&14 & 665  & 28 & 28  & 22 & -- & -- & 21\\
        H$_2$O &14 & 1085 & 38 & 46  & 32 & -- & -- & 43\\
        NH$_3$ &16 & 3608 & 123 & 130 & 125 & -- & -- & --\\
        N$_2$  &20 & 2950 & 72 & 76  & 71 & -- & -- & --\\
    \end{tabular}
    \end{ruledtabular}%
    }
    
    \label{table:results}
\end{table}

Table~\ref{table:results} reports the number of measurement groups obtained for the six Hamiltonians under full commutativity. Results for qubit-wise commutativity are not tabulated here because the number of groups is an order of magnitude larger and the model grows as $M\cdot K$. Thus, even for LiH, the QUBO would contain numbers of variables and couplers, $86940$ and $2,7\cdot 10^7$ respectively, (Table~\ref{tab:source}) far beyond the capacity of every backend considered, and the comparison would consist of a single row followed by a table filled with dashes. The reported quantity is $K_{\text{found}}$, the smallest number of groups for which a valid partition into mutually commuting sets was obtained. For the annealing backends, this is the smallest $K$ at which the descending loop of Section~\ref{sec:QUBO} still returned a zero-energy sample. It is an upper bound on the minimum rather than the minimum itself because failing to reach zero energy at $K-1$ does not prove infeasibility.

The ordering among the three classical colourings is not constant. LF is weakest on every instance, but the relative standing of RLF and DB changes with size. DB returns the smallest cover for LiH, BeH$_2$, H$_2$O and N$_2$ yet its margin reverses for NH$_3$ where it returns 125 groups against 123. This is consistent with the structure of the two algorithms, DB merges vertices according to their number of common neighbours and therefore exploits global information about the conflict graph, whereas LF fixes the colouring order once from the degree sequence and RLF rebuilds its candidate set at every colour. No single greedy, consequently, dominates the set.

Among the annealing backends, only the Leap Hybrid solver returned a cover beyond H$_2$. The two are limited for different reasons. Direct QPU sampling is limited by topology: from LiH onwards the source graph admits no minor embedding, as certified in Section~\ref{sec:Embedding}. This threshold is consistent with Ref.~\cite{Jattana2024}, whose annealer instances stop at a couple hundred logical variables. The H$_2$ instance under qubit-wise commutativity, at $M\cdot K = 70$ logical variables, is the largest molecular case common to both works, and the five groups reported there agree with the exact cover of Subsec.~\ref{sec:h2graph}. What Sec.~\ref{sec:Embedding} adds is the reason for the failure of every larger instance: it is not a sampling failure but the absence of any minor embedding, and it can be certified before any sample is drawn. Simulated annealing faces no embedding constraint, being a classical algorithm, but failed to return a zero-energy sample at any candidate $K$ within the compute budget allotted to it. The threshold, which lies between H$_2$ and LiH for direct QPU sampling, is the central limitation reported in this work and is analysed in detail in Section~\ref{sec:Embedding}.

Where the hybrid solver returned a cover, its quality can be compared with that of the best classical colouring. Table~\ref{table:results} shows that it performs better in two of the three cases: it reaches 20 groups for LiH, compared with 24 for DB, and 21 groups for BeH$_2$, compared with 22 for DB. For H$_2$O, however, the situation is reversed: the hybrid solver returns 43 groups, while DB returns 32, so the classical heuristic is clearly better on the largest of the three. The behaviour of the hybrid solver is not monotonic with instance size within this set. This is consistent with the model growing beyond the range in which the solver's internal classical heuristics remain effective, but three instances do not establish a trend and we claim none.

None of these partitions is close to certified optimality. Measured against the clique lower bounds of Table~\ref{tab:conflict}, the best cover found exceeds $\omega$ by a factor of $1.5$ for LiH, and $2$ for H$_2$O, rising up to $4.7$ for NH$_3$. Part of that growth reflects a looser bound rather than a worse partition, since the clique search uses fewer restarts on the larger graphs, but in every case the comparison between columns is a comparison between heuristics and not a measurement of the distance to the optimum.

Finally, the reduction achieved by grouping is substantial across the entire set. The ratio between the number of Pauli strings and the best number of groups found by any method lies between $29$ and $42$ for the five non-trivial instances, so a Hamiltonian with almost three thousand terms can be measured using about seventy distinct settings. This reduction makes the grouping problem worth solving, but it also makes the corresponding QUBO expensive because its size is the product of both quantities.

\subsection{Time to solution}

\begin{table*}[tp]
\caption{Cost of obtaining the partitions in Table~\ref{table:results} for the fully
commuting relation. The SA and QA entries are time-to-solution values~\cite{Ronnow2014} evaluated at the reported $K$ and include
annealing time only. The Hybrid column gives the accumulated cost of the full descending
loop over $K$, including the final infeasible value. A dash denotes an instance that the corresponding sampler could not
solve or embed. All results are obtained after a single execution of the algorithm.}
\begin{ruledtabular}
\begin{tabular}{l c | c c c | c c c}
\multirow{2}{*}{System} & \multirow{2}{*}{$M$}
& \multicolumn{3}{c|}{Classical (s)}
& \multicolumn{3}{c}{Annealing} \\
& & RLF & LF& DB 
& SA ($\mu$s) & QA ($\mu$s) & Hybrid (s) \\
\hline
H$_2$   &  14 &0.023 &0.008 &0.51 &512 &20 &2.98 \\
LiH & 630 &0.031 &0.019 &0.90 &-- &-- &892  \\
BeH$_2$ & 665 &0.040 &0.020 &1.12 &-- &-- &791  \\
H$_2$O & 1085 &0.115 & 0.035&4.83 &-- &-- &1220  \\
NH$_3$& 3608 &2.14 &0.145 &259 & --& --&--  \\
N$_2$& 2950 &2.23 &0.143 &165 &-- &-- & --\\
\end{tabular}
\end{ruledtabular}

\label{table:timetosolution}
\end{table*}

Table~\ref{table:timetosolution} shows that the rankings of the methods by quality and cost are not the same and that the spread in cost is far wider than the spread in quality. Across the entire set, the covers returned by the different methods differ by at most a factor of two, while their running times differ by more than three orders of magnitude.

Among the classical colourings, the ordering in time follows the complexity of the algorithms. LF is the cheapest for every instance, RLF is intermediate, and DB is the most expensive. The gap is not constant but widens sharply with instance size, from a factor of 47 over LF for LiH to 1790 for NH$_3$. Compared with LF, DB removes 31\%, 21\%, and 30\% of the groups for LiH, BeH$_2$, and H$_2$O, respectively, but only 7\% for N$_2$. For NH$_3$, it costs 120 times as much as RLF while returning a larger cover. The most expensive classical heuristic has therefore already passed the point of diminishing returns for the two largest instances in this set, which is precisely the scale at which an alternative method would need to be useful.

At that scale, the hybrid solver is expensive and its performance is inconsistent. For LiH, it spends 892~s, compared with 0.9~s for DB, a factor of about 990 to remove four groups. For BeH$_2$, it spends 791~s, compared with 1.12~s for DB, a factor of about 700 to remove one group. For H$_2$O, the trade-off fails in both directions: the hybrid solver spends 1220~s, compared with 4.83~s for DB, and returns a partition of 43 groups instead of 32. For the instances studied here, the hybrid solver is therefore not a practical method for solving the grouping problem, since it spends between a quarter and a third of an hour on a reduction that a classical heuristic either matches or surpasses in under five seconds.

It is important to specify precisely what the annealing column measures, because it is not the cost of solving the problem once. The QUBO of Eq.~(\ref{eq:QUBO_mvc}) encodes a decision problem at fixed $K$ and cannot be constructed until $K$ has been chosen, so the number of groups is an input to the model rather than an output. Obtaining the smallest $K$ requires the loop defined previously, which starts from the LF bound and incurs one full solver call for every candidate value, including the terminating call that, by construction, exhausts its budget without succeeding. For LiH, this sweep runs from $K=35$ down to the terminating call at $K=19$ and consists of seventeen calls averaging approximately 52~s each. H$_2$ is the only instance whose sweep consists of a single call, since the LF bound is already optimal and $K=1$ is rejected without sampling; a graph with at least one edge cannot admit a one-colour colouring.

This is a property of the formulation rather than of the hardware, and it applies to any solver used with this encoding. A greedy colouring returns the partition and its size together in one pass and needs no external estimate to begin. Only the accumulated sweep timing is comparable to the classical wall-clock.

The single case in which the quantum hardware is faster is H$_2$, where direct QPU sampling reaches the two-group partition with a time to solution of $20~\mu\text{s}$, compared with 8~ms for LF. This comparison should not be overinterpreted. The QPU figure counts annealing time alone; adding the programming and readout time of the call (91.8~ms) and the minor-embedding search (71~ms) brings the total to approximately 163~ms, roughly twenty times slower than the fastest classical method on the same instance.

Taken together with Table~\ref{table:results}, these results show that, for the system sizes studied here, the annealing-based approach is not cost-competitive by two to three orders of magnitude. Its advantage in partition size is confined to the two smallest non-trivial instances it can process and reverses for the third. The quantity of practical interest is therefore not a speedup but the boundary at which the formulation ceases to fit the hardware, as discussed in Section~\ref{sec:Embedding}.


\section{Conclusions}\label{sec:Conclusions}
We have formulated the Pauli-string grouping problem of the VQE as a standard one-hot colouring QUBO for the minimum clique cover of the commutativity graph, and studied its behaviour on a sparse-connectivity quantum annealer. This study covers six Hamiltonians in the STO-3G basis under both qubit-wise and full commutativity, evaluated on the Zephyr topology of the Advantage2 architecture and executed on the \texttt{Advantage2\_system1} processor.

The structural result is negative and quantitative. Of the twelve molecular instances, exactly two are embeddable on current annealing hardware, and both are H$_2$. Every other instance exceeds the $4\,800$ qubits of the ideal $Z_{12}$ graph by a factor between $23$ and $2.0\times10^{5}$, and the verdict is a certificate rather than a failed search. The degree-counting bounds of Eqs.~(\ref{eq:chain_bound}) and~(\ref{eq:qubit_bound}) settle it without running any embedding heuristic. Feeding those bounds back into the qubit and coupler counts of the ideal Zephyr family locates the requirement precisely. The smallest infeasible instance, BeH$_2$ under full commutativity, would need $Z_{59}$, more than $10^5$ qubits or the fabric of 25 Advantage2 processors. This requirement grows up to $Z_{5518}$ for NH$_3$ under qubit-wise commutativity. Because the source graphs are evaluated at the clique lower bound $\omega$ no admissible colour budget is smaller, so these are the most favourable figures the instances admit and no improvement in the colouring heuristics can move them.

The practical result is that this structural threshold is not the operative constraint, because a second limit is reached earlier. On the only instance benchmarked here, sampling costs about twenty times more wall clock than the cheapest classical colouring once programming readout and embedding search are counted, and it returns a partition that the same colouring obtains exactly and that the clique bound certifies as optimal before any sampling takes place. Where a hybrid solver bypasses the embedding problem entirely and does return smaller covers than the best classical heuristic, for LiH and BeH$_2$, it does so at two to three orders of magnitude more computation, and that advantage has already reversed by H$_2$O. There is therefore no size in this benchmark set at which annealing the grouping problem is preferable to colouring it classically. The threshold measured here does not delimit where a quantum annealer stops being useful for this task, it measures how far current hardware is from the size at which the question would become interesting.

Two of the obstacles belong to the formulation rather than to the hardware, and would persist on a larger processor. The first is that the model encodes a decision problem at fixed $K$. The number of groups is an input, so obtaining the smallest one requires an outer loop that pays a full solver call per candidate value, whereas a greedy colouring returns the partition and its size together in a single pass and needs no external estimate to begin. The second is the cost of the colour variables. Between $2\%$ and $38\%$ of the couplers of the source graph are the one-hot cliques of Eq.~(\ref{eq:source_counts}), which exist only to enforce feasibility and carry no information about the conflict graph, and the remainder is the conflict graph replicated once per colour. The encoding therefore multiplies the qubit floor imposed by $\tilde{G}$ alone by a factor of order $K$, between $13$ and $27$ under full commutativity and between $180$ and $1.6\times10^{3}$ under qubit-wise commutativity. Both are properties of the encoding and both would have to be addressed before additional connectivity could be exploited.

The leverage available is therefore of three kinds. The choice of commutativity relation is the largest, full commutativity needs between $2.5$ and $58$ fewer groups than qubit-wise commutativity on the same molecule, and through the quadratic dependence of the coupler count on $K$ this single factor separates the two relations by more than an order of magnitude. The representation of the Hamiltonian is the next. The same molecules expressed in symmetry-adapted orbitals yield appreciably fewer Pauli strings at identical Hartree-Fock energy, and since the model carries $M\cdot K$ variables, any reduction in $M$ propagates directly. Lastly it is worth studying the encoding itself. Leaner alternatives to the one-hot assignment, such as domain wall or binary encodings, kernelisation of the conflict graph before submission, or a formulation that places the number of groups in the objective rather than fixing it as a parameter, would reduce the variable count and, consequently, remove the outer loop. None of these is pursued here and quantifying them is the natural continuation of this work.

Finally, the scope of these conclusions should be stated plainly. They concern one architecture, one QUBO formulation, one basis set and one fermion-to-qubit mapping, and the classical baseline consists of three greedy colourings rather than an exact solver, so the distance to the true optimum is bounded only from below by a greedy clique estimate. What is architecture-independent is the counting, the source graph of the colouring QUBO is the Cartesian product $\tilde{G}\square K_K$, its size and degree sequence follow in closed form from $M$, $K$ and $|\tilde{E}|$, and the degree-counting bounds apply to any hardware graph of bounded degree. Those relations and the requirement curves they generate, remain valid as annealers grow.



\bibliographystyle{unsrtnat}
\hbadness=10000
\bibliography{references}

\end{document}